\documentclass[nojss]{jss}

\usepackage{amstext,amsfonts,amsmath,bm,thumbpdf,lmodern,hyperref}
\usepackage[all]{hypcap}

\usepackage{array,makecell,tikz,color,soul}
\usetikzlibrary{arrows.meta,positioning,shapes,arrows,decorations.pathreplacing,calc,automata,mindmap}

\newcommand{\argmin}{\operatorname{argmin}\displaylimits}
\newcommand{\argmax}{\operatorname{argmax}\displaylimits}

\newcommand{\I}{\mathbf{I}}

\definecolor{changes}{rgb}{1,1,0.7}

\numberwithin{equation}{section}

\title{Circular Regression Trees and Forests with an Application to Probabilistic Wind Direction Forecasting}
\Shorttitle{Circular Regression Trees and Forests}

\author{Moritz N. Lang\\Universit\"at Innsbruck
   \And Lisa Schlosser\\Universit\"at Innsbruck
   \And Torsten Hothorn\\Universit\"at Z\"urich
   \AND Georg~J.~Mayr\\Universit\"at Innsbruck
   \And Reto Stauffer\\Universit\"at Innsbruck
   \And Achim Zeileis\\Universit\"at Innsbruck}
\Plainauthor{Lang, Schlosser, Hothorn, Mayr, Stauffer, Zeileis}

\Abstract{
While circular data occur in a wide range of scientific fields, the methodology
for distributional modeling and probabilistic forecasting of circular response
variables is rather limited. Most of the existing methods are built on the
framework of generalized linear and additive models, which are often
challenging to optimize and to interpret. Therefore, we suggest circular
regression trees and random forests as an intuitive alternative approach that
is relatively easy to fit. Building on previous ideas for trees modeling
circular means, we suggest a distributional approach for both trees and forests
yielding probabilistic forecasts based on the von~Mises distribution. The
resulting tree-based models simplify the estimation process by using the
available covariates for partitioning the data into sufficiently homogeneous
subgroups so that a simple von~Mises distribution without further covariates
can be fitted to the circular response in each subgroup. These circular
regression trees are straightforward to interpret, can capture nonlinear
effects and interactions, and automatically select the relevant covariates that
are associated with either location and/or scale changes in the von~Mises
distribution. Combining an ensemble of circular regression trees to a circular
regression forest yields a local adaptive likelihood estimator for the
von~Mises distribution that can regularize and smooth the covariate effects.
The new methods are evaluated in a case study on probabilistic wind direction
forecasting at two Austrian airports, considering other common approaches as a
benchmark.
}


\Address{
  Moritz N. Lang, Lisa Schlosser, Achim Zeileis \\
  Universit\"at Innsbruck \\
  Department of Statistics \\
  Faculty of Economics and Statistics \\
  Universit\"atsstr.~15 \\
  6020 Innsbruck, Austria \\
  E-mail: \email{Moritz.Lang@uibk.ac.at}, \\
  \phantom{E-mail: }\email{Lisa.Schlosser@uibk.ac.at},\\
  \phantom{E-mail: }\email{Achim.Zeileis@R-project.org} \\
  URL: \url{https://www.uibk.ac.at/statistics/personal/moritz-lang/},\\
  \phantom{URL: }\url{https://www.uibk.ac.at/statistics/personal/schlosser-lisa/},\\
  \phantom{URL: }\url{https://eeecon.uibk.ac.at/~zeileis/}\\

  Reto Stauffer \\
  Universit\"at Innsbruck \\
  Digital Science Center and \\ 
  Department of Statistics \\
  Faculty of Economics and Statistics \\
  Universit\"atsstr.~15 \\
  6020 Innsbruck, Austria \\
  E-mail: \email{Reto.Stauffer@uibk.ac.at} \\
  URL: \url{https://retostauffer.org/}\\
  
  Georg~J.~Mayr\\
  Universit\"at Innsbruck \\
  Department of Atmospheric and Cryospheric Science \\
  Faculty of Geo- and Atmospheric Sciences \\
  Innrain~52f \\
  6020 Innsbruck, Austria \\
  E-mail: \email{Georg.Mayr@uibk.ac.at} \\
  URL: \url{https://www.uibk.ac.at/acinn/people/georg-mayr}\\
  
  Torsten Hothorn\\
  Universit\"at Z\"urich \\
  Institut f\"ur Epidemiologie, Biostatistik und Pr\"avention \\
  Hirschengraben 84\\
  CH-8001 Z\"urich, Switzerland \\
  E-mail: \email{Torsten.Hothorn@R-project.org}\\
  URL: \url{http://user.math.uzh.ch/hothorn/}\\
  
}

\begin{document}

\section{Introduction}
\label{sec:introduction}

Circular data can be found in a variety of applications and subject areas,
e.g., hourly crime rate in socio-economics, animal movement direction or
gene-structure in biology, and wind direction as one of the most important
weather variables in meteorology. Fitting a statistical model to this type of
data requires the incorporation of its specific feature of periodicity. For
example, angular data are restricted to an interval such as~$[0,2\pi)$ with~$0$
being equivalent~to~$2\pi$.

\subsection[Circular regression: Conditional mean vs. distributional models]{Circular regression: Conditional mean vs.\ distributional models}

Many approaches to model circular data assume that the circular variable of
interest follows a circular distribution, in particular the von~Mises
distribution which is also known as the ``circular normal distribution''. One
of the first regression models with a circular response variable and linear
covariates was presented by \cite{Gould:1969} where the circular mean is
predicted by a linear combination of covariates. \cite{Johnson+Wehrly:1978}
refined this idea by plugging in a link function transforming the linear
predictor to a restricted interval of length~$2\pi$. This generalized linear
model (GLM) type approach was further extended by \cite{Fisher+Lee:1992} and
subsequently by \cite{Fisher:1993} introducing independent GLMs for either the
location or scale parameter with appropriate link functions while keeping the
other parameter constant. Additionally, they developed a combined
heteroscedastic or distributional version with alternating reestimation of the
location and scale parameters conditional on the respective sets of covariates
until convergence. While all of these models are built on well-elaborated
theory, their application in practice remains very challenging, mainly due to
the complexity encountered in optimizing the corresponding log-likelihood
function which is not globally concave. Therefore, highly-informative starting
values are crucial for such circular GLMs to converge
\citep{Pewsey+Neuhaeuser+Ruxton:2013, Gill+Hangartner:2010}. In order to avoid
this strong dependence on appropriate initial values,
\cite{Mulder+Klugkist:2017} present a Bayesian alternative of a homoscedastic
GLM for circular data. However, apart from potential difficulties in the
optimization procedure of circular GLMs, the interpretation of the underlying
additive effects is often challenging as well because the link function is
highly nonlinear and the representation of smooth transitions on the unit
circle is not straightforward. For example, the same rotation can be obtained
in either positive or negative direction on the circle leading to an ambiguous
interpretation.

As a very intuitive and data-driven alternative, we propose a flexible
tree-based regression approach for modeling circular data by applying the
von~Mises distribution within the methodology of distributional trees and
forests \citep{Schlosser+Hothorn+Stauffer:2019}. The resulting circular
regression trees and forests avoid the discussed difficulties of circular GLMs
by using the available covariates for partitioning the data into sufficiently
homogeneous subgroups so that a simple von~Mises distribution without further
covariates can be fitted to the circular response in each of these subgroups.
This obviates the need for a link function or for iterating between models for
the separate distribution parameters. By leveraging the distributional modeling
approach, the trees can automatically detect and capture differences in both
distribution parameters, providing a fully specified circular response
distribution in each terminal node offering a wide range of statistical
inference. In addition, the employed tree structure allows to capture
non-additive effects while forests enable the modeling of smooth changes.
Furthermore, covariates and their possible interactions do not need to be
specified in advance as they are selected automatically in the recursive
partitioning algorithm.

This novel approach to circular regression trees and forests complements the
literature on tree-based circular modeling. \cite{Lund:2002} already introduced
a circular regression tree algorithm where binary splits are made based on an
angular distance measure capturing node homogeneity. However, this only models
changes in the conditional mean but not the conditional variance or full
probabilistic distribution which would allow for also considering uncertainty
in forecasts \citep{Gneiting:2008}. In contrast, we introduce a distributional
tree approach that considers splits in all distribution parameters and yields a
full probabilistic predictive model. In addition, as a natural extension,
ensembles or forests of circular regression trees are presented in this study,
showing that these can further improve predictive power.

\begin{figure}[p!]
\centering
\includegraphics[width=0.9\textheight,,angle=90,origin=c]{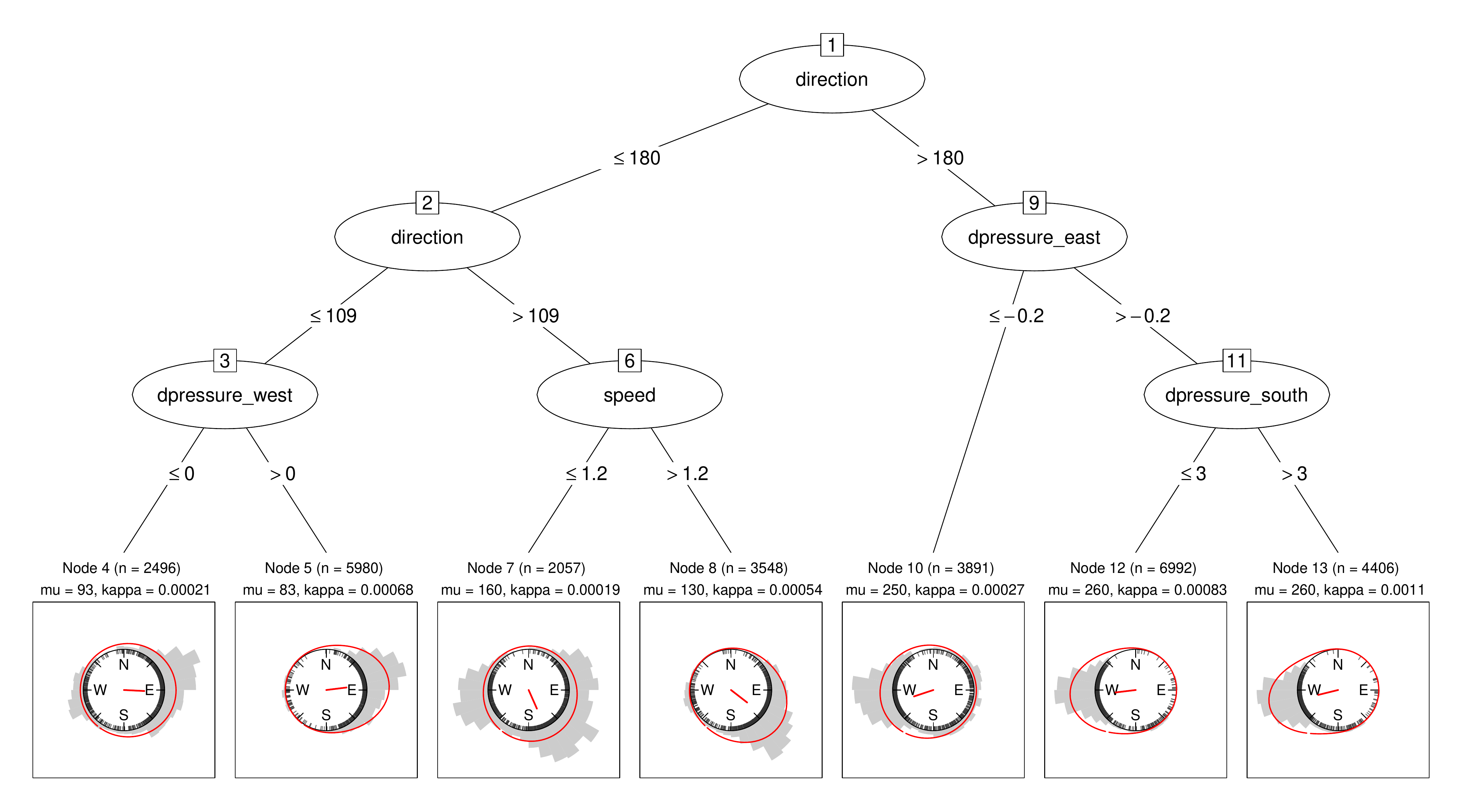}
\caption{\label{fig:tree_ibk} Fitted tree based on the von~Mises distribution
for 1-hourly wind direction forecasts at the airport of Innsbruck. In each
terminal node the empirical histogram (gray) and fitted density (red line) are
depicted along with the estimated location parameter (red hand). The covariates
selected for splitting are wind direction (meteorological degree), wind speed
($\text{ms}^{-1}$), and pressure gradients ($\text{dpressure; hPa}$) west, east
and south of the airport, all lagged by one hour. Note that in the
meteorological context wind direction is defined on the scale $[0,360]$~degree
and increases clockwise from North ($0$ degree).}
\end{figure}

\subsection{Motivating example}

To provide a first impression of the presented methodology, a circular
regression tree is employed for probabilistic wind direction forecasting. Wind
direction is a classical circular quantity and accurate forecasts are of great
importance for decision-making processes, e.g., in air traffic management as
considered in this study. Figure~\ref{fig:tree_ibk} shows an estimated tree for
1-hourly forecasts at Innsbruck Airport, located at the bottom of a narrow
valley within the European Alps. Topography channels the wind along the
west-east valley axis or along a tributary valley intersecting from the south.
Hence, pressure gradients to which valley wind regimes react are considered as
covariates along with other meteorological measurements (lagged by one hour)
and their derivatives, such as wind direction and wind speed at the airport
itself as well as spatial and temporal differences.

Figure~\ref{fig:tree_ibk} illustrates the resulting tree along with the
empirical (gray) and fitted von~Mises~(red) wind direction distribution in each
terminal node. Based on the fitted location parameters~$\hat \mu$, the
subgroups can be distinguished into the following wind regimes: (1)~Up-valley
winds blowing from the valley mouth towards the upper valley (from east to
west, nodes~4 and 5); (2)~Downslope winds blowing across the Alpine crest along
the intersecting valley towards Innsbruck (from south-east to north-west,
node~8); (3)~Down-valley winds blowing in the direction of the valley mouth
(from west to east, nodes 10, 12 and~13). Node~7 captures observations with
rather low wind speeds that cannot be clearly distinguished into specific wind
regimes and are consequently associated with a very low estimated concentration
parameter~$\hat{\kappa}$, i.e., a high estimated variance. In terms of
covariates, the lagged wind direction (``persistence'') is mostly responsible
for distinguishing the broad range of wind regimes listed above while the
pressure gradients and wind speed separate the data into subgroups with high
vs.\ low precision. A more extensive case study of circular regression trees
and forests applied to probabilistic wind direction forecasting at Innsbruck
Airport and Vienna International Airport is presented in
Section~\ref{sec:wind}, along with a benchmark against commonly-used
alternative approaches.

The remainder of the paper is structured as follows: The theory on
probabilistic circular modeling introducing the von~Mises distribution, and
circular regression models are discussed in Section~\ref{sec:prob_circ}. The
methodology of circular regression trees and forests and their features are
introduced in Section~\ref{sec:tree_forest}. After the case study presented in
Section~\ref{sec:wind}, a comprehensive summary and conclusions are given in
Section~\ref{sec:summary}.

\section{Probabilistic circular modeling}
\label{sec:prob_circ}
Probabilistic modeling of circular data requires the selection of a probability
distribution which accounts for the periodicity of circular data. Generally,
this feature can be obtained by ``wrapping'' the probability density function
of any continuous distribution around the unit circle \citep{Mardia+Jupp:1999}.
In that way, the wrapped Cauchy distribution or the wrapped normal distribution
can be employed to model symmetric unimodal circular data. A close
approximation to the wrapped normal distribution that is mathematically simpler
and hence easier to use is provided by the von~Mises distribution
\citep{Fisher:1993}, a purely circular distribution which is also known as
``the circular normal distribution'' and is a common choice for probabilistic
modeling of circular data. Based on a location parameter $\mu \in [0, 2\pi)$
and a concentration parameter $\kappa > 0$ the density of the von~Mises
distribution for an observation $y \in [0, 2\pi)$ is given by
\begin{equation}
  f_\mathrm{vM}(y; \mu, \kappa) = \frac{1}{2 \pi I_0(\kappa)}~e^{ \kappa \cos(y - \mu)},\label{schlosser:equ_vm}
\end{equation}
where $I_0(\kappa)$ is the modified Bessel function of the first kind and order
$0$ (see, e.g., \citealt{Mardia+Zemroch:1975b}, \citealt{Jammalamadaka+Sengupta:2001}, or
\citealt{Ley+Verdebout:2017} for a more detailed overview).

\begin{figure}[t]
  \begin{minipage}{.05\textwidth}
    \hfill
  \end{minipage}%
  \begin{minipage}{.45\textwidth}
    \vspace{-1em}
    \includegraphics[width=1.05\linewidth]{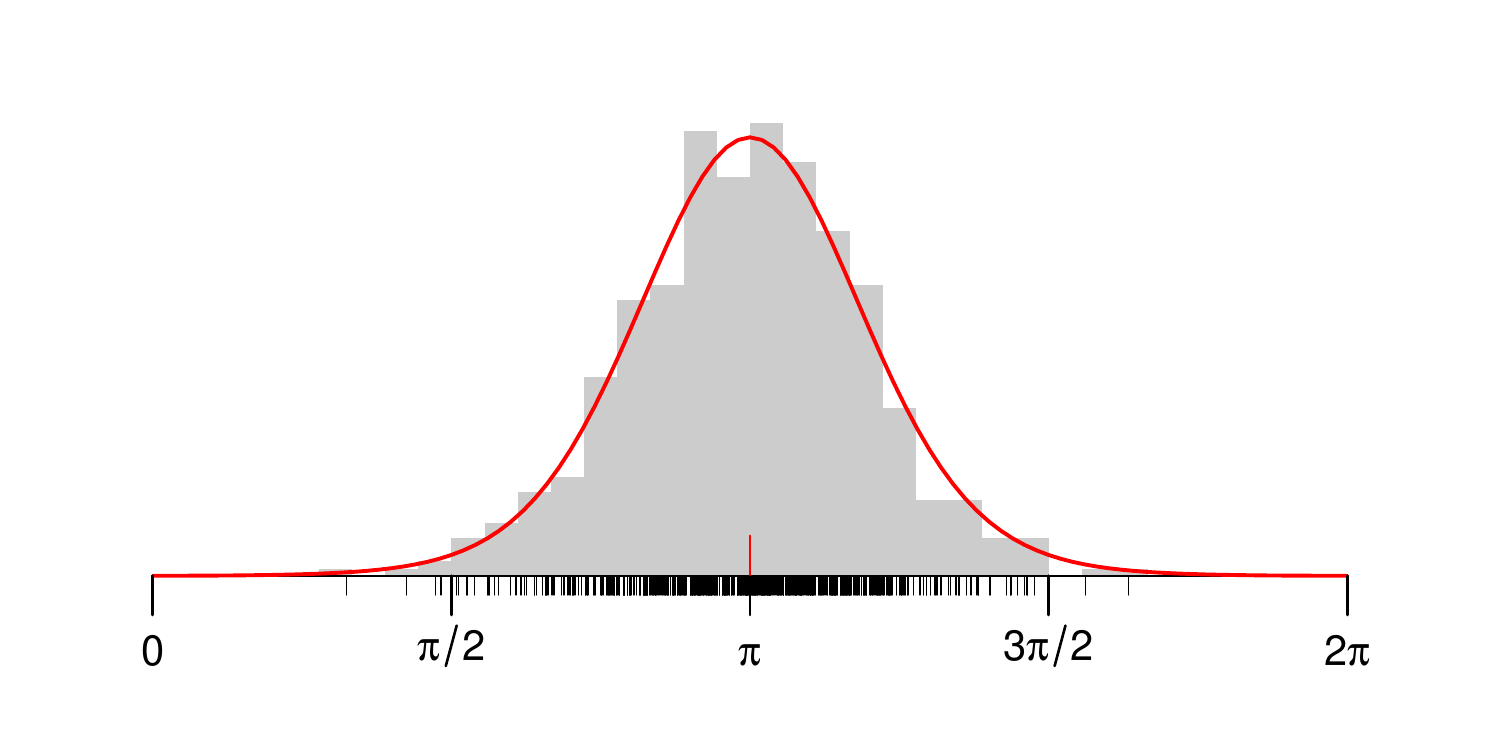}
  \end{minipage}%
  \begin{minipage}{.45\textwidth}
    \includegraphics[width=1.2\linewidth]{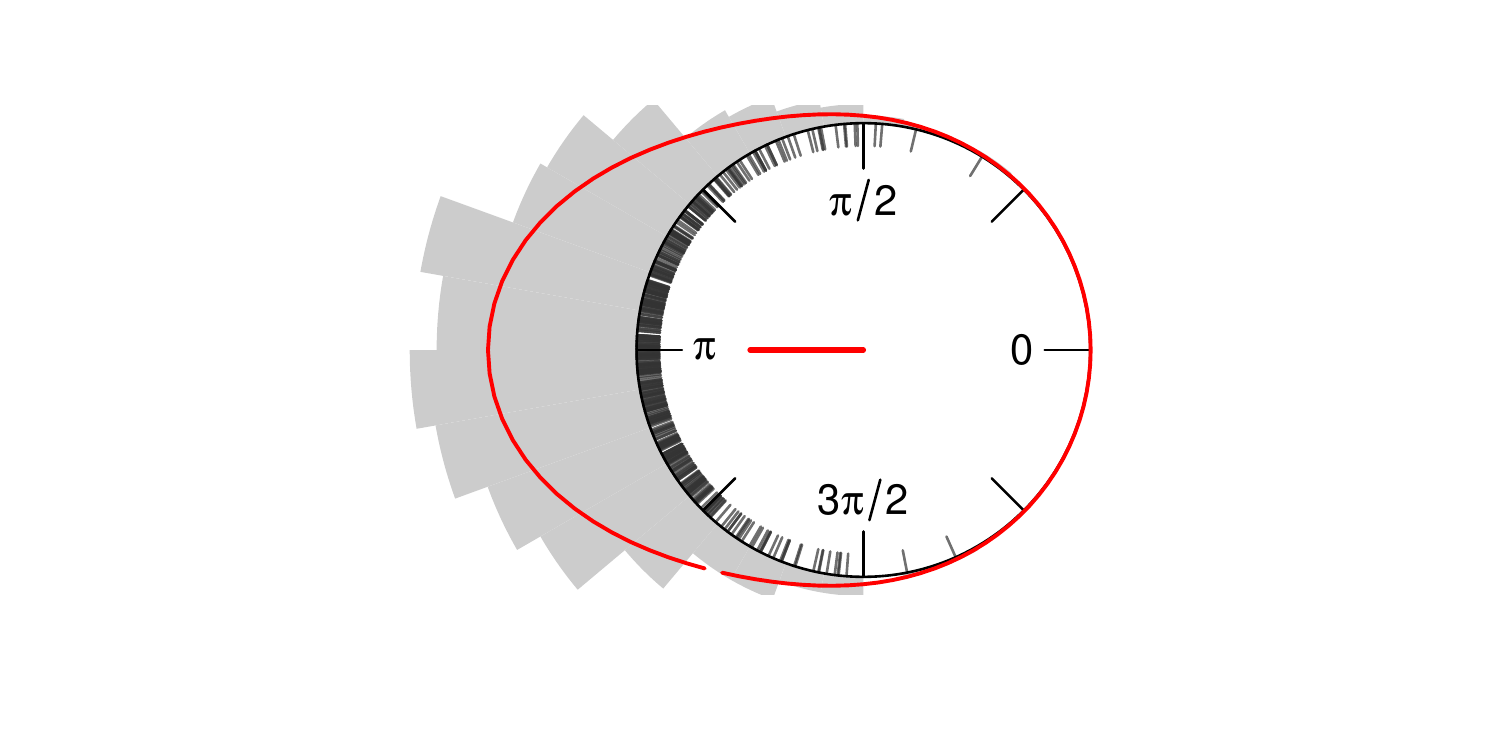}
  \end{minipage}
  \begin{minipage}{.05\textwidth}
    \hfill
  \end{minipage}%
  \caption{\label{fig:densities} Illustration of a von~Mises model for circular data in
    the interval $[0, 2\pi)$ fitted by maximum likelihood. Left: Linearized
    scale. Right: Circular scale. In both panels the empirical histogram (gray bars)
    and fitted density (red line) are depicted along with the estimated location
    parameter (red hand).}
\end{figure}

The corresponding log-likelihood function is defined as
\begin{equation}
    \ell(\mu, \kappa; y) = \log(f_\mathrm{vM}(y;\mu, \kappa))
    = -\log(2 \pi I_0(\kappa)) + \kappa \cos(y - \mu).\label{equ:vM:loglik}
\end{equation}
To fit a probabilistic model $\text{vM}(y; \mu, \kappa)$ to a circular response
$y$, the distribution parameters $\mu$ and $\kappa$ need to be estimated. This
can be done by maximizing the log-likelihood function $\ell(\mu, \kappa; y)$
\begin{equation}\label{eq:optim}
  (\hat{\mu},\hat{\kappa}) = \argmax_{\mu,\kappa} \sum_{i=1}^n \ell(\mu, \kappa; y_i) 
\end{equation}
yielding maximum likelihood estimators $\hat{\mu}$ and $\hat{\kappa}$ such that
a fully specified distributional model is fitted to the learning data
$\{y_i\}_{i=1,\ldots,n}$.

The score function
\begin{equation}
  \label{eq:scores}
  \begin{aligned}
    s(\mu, \kappa, y) &= \bigg(\frac{\partial \ell}{\partial \mu}(\mu,\kappa; y)
    \smallskip &,
    \quad & \frac{\partial \ell}{\partial \kappa}(\mu,\kappa; y) \bigg) 
    \\
    &= \bigg(\kappa \sin(y-\mu)
    \smallskip &,
    \quad &-\frac{I_1(\kappa)}{2\pi I_0(\kappa)}+\cos(y-\mu) \bigg)
  \end{aligned}
\end{equation}
provides a way to obtain a measure of goodness of fit of the model for each
observation and fitted parameter. Then, the optimization problem in
Equation~\ref{eq:optim} can alternatively be specified as 
\begin{equation}
  \sum_{i = 1}^n s(\hat{\mu}, \hat{\kappa}, y_i) = 0.
\end{equation}
Figure~\ref{fig:densities} depicts a von~Mises model for circular data in $[0, 2
\pi)$ fitted by maximum likelihood, either using a linearized (left) or
circular (right) scale. In both cases, the empirical histogram (gray bars) is
shown along with the fitted density (red line) and estimated location parameter
(red hand). However, this distributional model only considers the circular
response variable but no covariate. Of course, including covariates is of
interest in a regression setup for forecasting.

In most generalized linear model (GLM) or generalized additive model (GAM)
approaches to circular regression the location parameter $\mu$ depends on
covariates $\mathbf{z}$ through a link function~$g(\cdot)$, circular intercept
$\mu_0$ and coefficient vector $\mathbf{\beta}$: \begin{equation} \mu = \mu_0 +
g\left(\mathbf{\beta}^{\top} \mathbf{z}\right). \end{equation} The link
function transforms the additive predictor to an interval of length $2\pi$.
Typically $g(x) = 2\cdot \text{arctan}(x)$ is employed, as suggested by
\citet{Fisher+Lee:1992}. They also developed a heteroscedastic version by
combining two individual GLMs, each for one of the parameters~$\mu$ and~$\kappa$. 
This provides a first approach to a fully probabilistic regression
model for circular data, albeit the parameters are not regressed simultaneously
on covariates as in the more general framework provided by generalized additive
models for location, scale, and shape \citep[GAMLSS,
][]{Rigby+Stasinopoulos:2005}. Nevertheless, other circular additive models
share the previously discussed difficulties induced by the characteristics of
the log-likelihood function and the strongly nonlinear link function. Referring
to additive models in general, it has to be considered that a proper model
specification can be very challenging, particularly for a high number of
covariates and no information on possible interactions. Moreover, the additive
structure might impose a smooth effect even if the true underlying effect is an
abrupt shift, which occurs, e.g., in atmospheric wind fields.

In contrast, the tree-based circular regression models proposed in the next
sections largely avoid the problems above by employing recursive partitioning
in combination with local adaptive likelihood estimation.

\section{Circular regression trees and forests}
\label{sec:tree_forest}

Starting out from the ideas of \citet{Lund:2002}, we introduce circular
regression trees and forests considering splits in all distribution parameters
of the von~Mises distribution and providing a full probabilistic model.
Moreover, the resulting tree-based models provide a very intuitive and
data-driven alternative to commonly used GLMs for circular data.

\subsection{Circular regression trees}
\label{sec:circtree}
Fitting a global model to a full data set can be very challenging, particularly
for complex data with substantial variations. Therefore, separating the data
set into more homogeneous subgroups based on covariates before fitting a local
model in each of these subgroups allows to capture (potential) group-specific
effects more precisely and hence can result in an overall better-fitting model.
This is the general idea of regression trees which are combined with
distributional modeling in \cite{Schlosser+Hothorn+Stauffer:2019}. Specifying a
full distributional model in each node of the tree yields a distributional
regression tree, where selecting the von~Mises distribution enables an
application to circular data. The crucial step of how and where to split the
data can be accomplished with the unbiased recursive partitioning algorithms
MOB \citep{Zeileis+Hothorn+Hornik:2008} or CTree
\citep{Hothorn+Hornik+Zeileis:2006}. For this purpose, model scores are
obtained by evaluating the score function~$s(\cdot)$ for each individual
observation at the parameter estimates (Equation~\ref{eq:scores}). For the
von~Mises distribution with its two distribution parameters ($\mu$ and
$\kappa$) and a data set of $n$ observations, this yields an $n \times 2$
matrix that can be employed as a discrepancy measure, capturing how well each
given observation conforms with the estimated location $\hat{\mu}$ and
precision $\hat{\kappa}$, respectively. To capture dependence on covariates,
the association between the model's scores and each available covariate is
assessed using either a parameter instability test~(MOB) or a permutation
test~(CTree). By doing so in each partitioning step, the covariate with the
highest significant association (i.e., lowest significant $p$-value, if any) is
selected for splitting the data. The corresponding split point is chosen either
by optimizing the log-likelihood~(MOB) or a two-sample test statistic~(CTree)
over all possible partitions. This procedure is repeated recursively until
there are no significant parameter instabilities or until another stopping
criterion is met (e.g., subgroup size or tree depth). A more detailed
description of the applied tree-building algorithm can be found in
Appendix~\ref{app:algorithm}.

Once a distributional tree model is fitted it can be employed to obtain
probabilistic predictions for a possibly new set of observed covariates $\bm{z}
= (z_1, \ldots, z_m)$. Starting at the root node, the tree structure leads the
observation to a terminal node where the parameter pair $(\hat{\mu},
\hat{\kappa})$ is estimated for the corresponding subset of learning
observations. This can also be expressed by employing weights which indicate
whether the $i$-th learning observation and the observation~$\bm{z}$ belong to
the same terminal node: 
\begin{equation}
  w^{\text{tree}}_i(\bm{z}) = \sum_{b=1}^B \mathbf{1}((\bm{z}_i \in
  \mathcal{B}_b) \land (\bm{z} \in \mathcal{B}_b)).
\end{equation} 
Here, $\mathbf{1}(\cdot)$ is the indicator function and $\mathcal{B}_b$ is the
$b$-th out of $B$ segments partitioning the covariate space in disjoint
subsets. Then the estimated parameter pair $(\hat{\mu},\hat{\kappa})(\bm{z})$
specifying the predicted von~Mises distribution for a given $\bm{z}$ is
obtained by a weighted maximum likelihood estimator:
\begin{equation} 
  (\hat{\mu},\hat{\kappa})(\bm{z}) =
  \argmax_{\mu,\kappa} \sum_{i=1}^n w^{\text{tree}}_i(\bm{z}) \cdot
  \ell(\mu,\kappa; y_i).
\end{equation}
Therefore, the same parameter pair is estimated for all observations belonging
to the same terminal node, which speeds up computation since the parameter
estimates do not need to be recalculated for each (new) observation via maximum
likelihood but can be extracted directly from the learning sample and the
fitted model.

While tree models can capture non-additive effects, their structure and the
consequential strict separation of data into subgroups hinders an adequate
depiction of smooth effects. They can be included by combining an ensemble of
trees in order to obtain a regression forest, which also stabilizes the model.

\subsection{Circular regression forests}
\label{sec:circforest} 
A natural extension of (circular) regression trees are ensembles or forests
that can improve forecasts by regularizing and stabilizing the model. Random
forests introduced by \cite{Breiman:2001} average the predictions of an
ensemble of trees, each built on a subsample or bootstrap of the original data.
A generalization of this strategy is to obtain weighted predictions by adaptive
local likelihood estimation of the distributional parameters
\citep[Section~2.3. of][]{Schlosser+Hothorn+Stauffer:2019,
Hothorn+Zeileis:2017}. More specifically, for each (possibly new)
observation~$\bm{z}$ a set of averaged ``nearest neighbor'' weights
$w_i^\text{forest}(\bm{z})$ is obtained that is based on the number of trees in
which $\bm{z}$ is assigned to the same terminal node as each learning
observation $y_i, i \in \{1,\ldots,n\}$. Hence, for a forest of $T$ trees, the
weights are calculated via
\begin{equation}
  w^{\text{forest}}_i(\bm{z}) = \frac{1}{T} \sum_{t=1}^T \sum_{b=1}^{B^t}
  \frac{\mathbf{1}((\bm{z}_i \in \mathcal{B}^t_b) \land (\bm{z} \in \mathcal{B}^t_b))}{|\mathcal{B}^t_b|},
\end{equation}
where $|\mathcal{B}^t_b|$ denotes the number of observations in the $b$-th
segment of the $t$-th tree. Therefore, similar observations ending up more
often in the same terminal node have higher weights and thus a stronger
influence in the weighted maximum likelihood estimation.

In that way a specific set of weights can be calculated for each observation
yielding its specific parameter estimates for the von~Mises distribution
\begin{equation}
  (\hat{\mu},\hat{\kappa})(\bm{z}) = \operatorname{argmax}\displaylimits_{\mu,
  \kappa} \sum_{i=1}^n w_i^\text{forest}(z) \cdot \ell(\mu, \kappa; y_i). 
\end{equation}

Therefore, the resulting parameter estimates can smoothly adapt to the given
covariates $\bm{z}$ whereas $w_i(\bm{z}) = 1$ would correspond to the
unweighted full-sample estimates and $w_i(\bm{z}) \in \{0, 1\}$ corresponds to
the subgroup selection from a tree. Thus, circular regression forests can
capture both smooth and abrupt changes, while covariates and possible
interactions are selected automatically and do not explicitly need to be
specified beforehand.

\section{Case study: Probabilistic wind direction forecasting}
\label{sec:wind}
As motivated in Section~\ref{sec:introduction}, accurate forecasts of wind
directions are of great importance for risk management in various fields such
as agriculture, energy production, or aviation. For example, in order to direct
airplanes to a safe landing, precise knowledge of wind direction for the next
hour(s) at the respective airport is highly desirable and adequate prediction
methods are required. This section exemplifies the use of circular regression
trees and forests with wind direction forecasts for two Austrian airports --
one in flat terrain, the other one in mountainous terrain. The results are
benchmarked against alternative probabilistic forecasting methods. The study is
based on $+1$\,h and $+3$\,h forecasts employing lagged observations in the
vicinity of the airports as possible predictor variables. 

\subsection{Data}
\begin{figure}[t]
  \setkeys{Gin}{width=1.0\textwidth}
  \centering
  \includegraphics{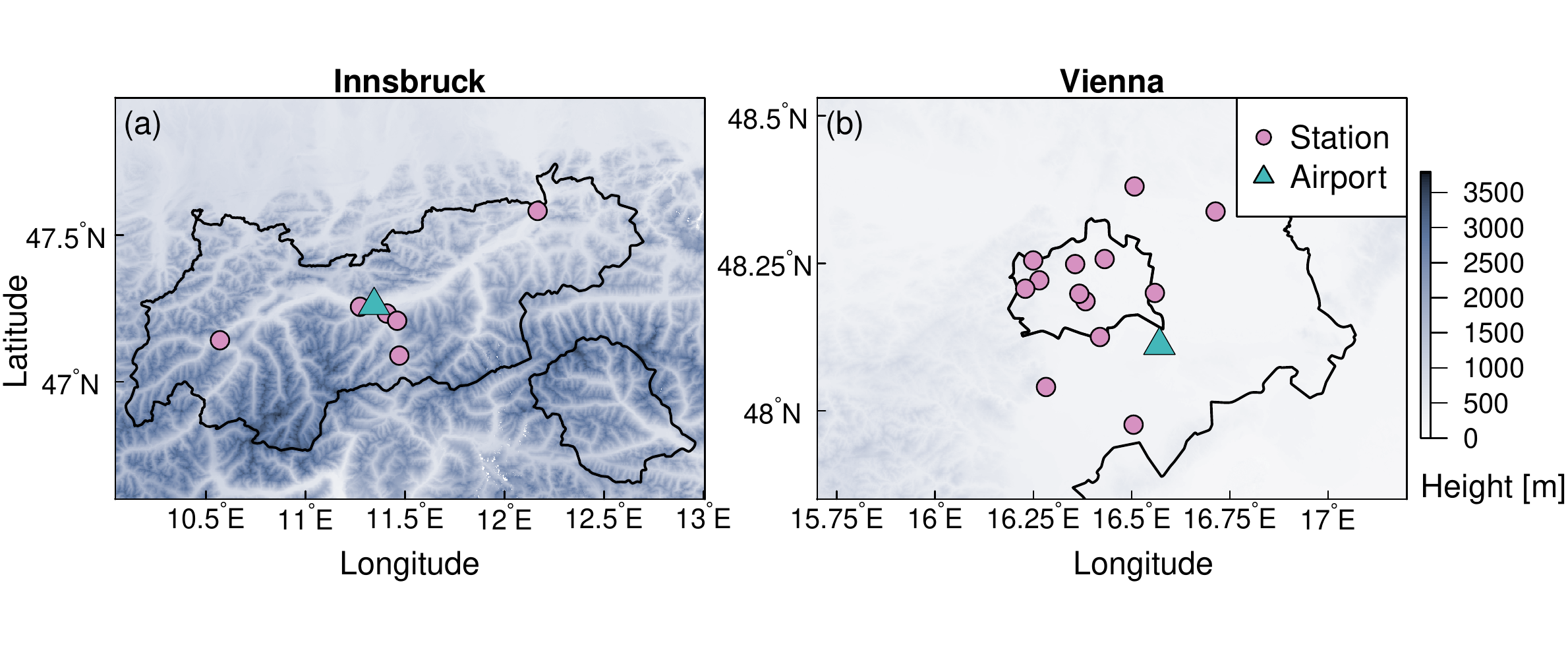}
  \caption{Overview of the study area for Innsbruck Airport (left) and Vienna International Airport (right). 
    For Innsbruck, $4$~stations at the airport and $6$~stations along the
    intersecting valleys are used, whereas, for Vienna, $9$~stations at the airport
    and $13$~stations within the vicinity of approximately 30\,km are used.
    Elevation data are obtained from the TandDEM-X digital elevation model with a 
    horizontal resolution of $90$~m \citep{Wessel:2018}.}
  \label{fig:overview} 
\end{figure}
\label{sec:wind:data}
The circular response variable considered in this case study is a
$10$\,min-average of wind direction measurements at Innsbruck Airport (INN) and
Vienna International Airport (VIE) on an hourly temporal resolution. Temporal
information and $1$-hourly resolved $10$\,min mean observations of various
meteorological quantities are used as predictor variables, including wind
direction, wind speed, temperature, air pressure and humidity, all lagged by
one or three hours according to the respective forecasting step. The
meteorological variables are measured either directly at the airports or within
their vicinities. For Innsbruck, measurements at the airport and along the
intersecting valleys are used, whereas, for Vienna, measurements at the airport
and within its vicinity of approximately 30\,km are used.
Figure~\ref{fig:overview} provides a topographical overview of the airports and
their surrounding areas with the station sites employed in this study. In
addition, we use derived quantities such as 3-hourly means, minima and maxima,
as well as 1-and 3-hourly temporal changes and spatial differences towards the
airport of the respective quantities. An overview of the employed data sets can
be found in Table~\ref{tab:data}. 

The data used in this study consists of five subsequent years from January~2014
to December~2018. After first eliminating predictor variables with more than
5\% missing values and then time points with any missing observations, the data
set consists of 41\,979~time points and $260$~covariates for Innsbruck, and of
38\,985~time points and $494$~covariates for Vienna, respectively.

\begin{table}[t!]
 \caption[Table caption text]{Overview of the data sets employed in the case
    study: For Innsbruck and Vienna, various meteorological variables and derived
    quantities of these are considered at the respective stations, located either
    directly at the airports or in their vicinities.}
  \label{tab:data}
  \begin{center}
    \begin{tabular}{l  l}
      \hline
      Data components               & Description \\
      \hline
      Temporal information:         & Time of the day, day of the year \\
      \noalign{\vskip 1mm}   
      Meteorological variables:     & Wind direction, wind (gust) speed, \\
                                    & (reduced) air pressure, relative humidity, \\
                                    & temperature\\ 
      \noalign{\vskip 1mm}                           
      Derived quantities:           & 3-hourly means/minima/maxima, \\
                                    & 1-hourly and 3-hourly temporal changes, \\
                                    & spatial differences towards the airport\\
      \noalign{\vskip 1mm}                              
      Weather stations (Innsbruck): & 4 stations at the airport, as well as\\
                                    & Igls, Kematen, Kufstein, Landeck, Patscherkofel, \\
                                    & and Steinach\\
      \noalign{\vskip 1mm}
      Weather stations (Vienna):    & 9 stations at the airport, as well as\\
                                    & Arsenal, Donaufeld, Exelberg, G\"anserndorf, \\
                                    & Gro{\ss}-Enzersdorf, Gumpoldskirchen, Hohe Warte,\\
                                    & Innere Stadt, Jubil\"aumswarte, Mariabrunn, \\
                                    & Seibersdorf, Unterlaa, and Wolkersdorf\\
      \hline
    \end{tabular}
  \end{center}
\end{table}

\subsection{Models and evaluation}
\label{sec:wind:models}
For a fair evaluation of circular regression trees and forests, and to
investigate whether they can be applied as a reasonable alternative to already
existing approaches, three additional statistical models are employed in this
study for probabilistic forecasting of wind directions. Two of them are based
on existing approaches used in the meteorological field, while the third is a
state-of-the-art GLM-type model to forecast circular response variables.

\begin{itemize}
  \item \emph{Climatological model:} 
    Accurate knowledge of weather quantities' climatologies can be important for a
    wide range of applications. While forecasts based on climatologies, by
    construction, do not adapt to the current weather situation they are still a
    useful baseline for the validation of newly developed forecasting systems
    \citep{Simon+Umlauf+Zeileis:2017, Stauffer+Mayr+Messner:2017}. 
    \newline
    Specifically, the climatology employed in the following uses all observations
    at the same time (to adapt to daily cycles) in a window of $31$~days centered
    around the day of interest (to adapt to seasonal cycles)
    in all available years in the sample. Based on these observations a probabilistic
    model is obtained by maximum likelihood estimation as described in
    Section~\ref{sec:prob_circ}. This approach follows \citet{Vogel+Knippertz+Fink:2018}
    and is discussed in a comprehensive summary on different time-adaptive training schemes in
    \citet{Lang+Lerch+Mayr:2019}.
  \item \emph{Persistency model:}
    The persistence describes the previous value of a single weather quantity
    in a time series. Like the climatology it is a very basic prediction model that
    is often applied as a baseline reference in weather forecasts
    \citep{noaasnationalweatherservice:2019}. Especially in nowcasting tasks with very short
    forecasting steps, the persistence can provide very good estimates.
    \newline
    To gain a full probabilistic persistency model, we proceed similarly as for
    the climatological model by using maximum likelihood estimation and fitting the
    distribution parameters of the von~Mises distribution conditional on lagged
    response values according to the description in Section~\ref{sec:prob_circ}.
    We fit one model for every hour throughout the validation data set employing
    the previous six lagged response values as training data. In order to allow for
    a stronger influence of observations closer to the time of interest,
    exponential smoothing is employed with a smoothing factor of $0.5$;
    accordingly, for every prediction an equal influence rate of 50 percent is
    assigned both to the current observation and to the previous five observations
    together. Observations with longer time lags have exponential weights below
    $0.01$ and are therefore omitted from the training data.
  \item \emph{Generalized linear model:}
    Traditional approaches to forecast circular response variables are often
    based on circular GLM-type models \citep{Fisher:1993}. As discussed in
    Section~\ref{sec:introduction}, circular regression models often experience the
    problem that the likelihood function can be strongly irregular which makes
    optimization rather difficult. Hence, they often do not converge if no
    appropriate initial values are provided \citep{Pewsey+Neuhaeuser+Ruxton:2013,
    Gill+Hangartner:2010}. In this study, to be able to employ a GLM out of the box
    as a reference, we use the Bayesian implementation of
    \cite{Mulder+Klugkist:2017} which depends less on initial values due to an MCMC
    sampling algorithm using weakly informative priors.  \newline Following
    \cite{Mulder+Klugkist:2017} the model uses a link function~$g(\cdot)$ to keep the
    response values within an interval of length $2\pi$. As the implementation
    cannot handle circular covariates, we use the components of the lagged
    2-dimensional wind vector~$(u, v)^{\top}$ and the lagged
    wind speed $spd$ as predictor variables. The model formula for the location
    parameter $\mu$ of the von~Mises distribution can be written as:
    \begin{align}
      \mu = \beta_0 + g(\beta_1 \cdot u + \beta_2 \cdot v + \beta_3 \cdot spd)
    \end{align}
    with $\beta_0$ being a circular intercept, $\beta_\bullet$ the regression
    coefficients and the link function $g(x) = 2 \cdot \text{arctan}(x)$. In
    addition, a constant concentration parameter $\kappa$ is fitted to the full
    learning sample. To allow for seasonally varying error characteristics in both
    bias and slope coefficients, and to allow for seasonal heteroscedasticity
    captured by the concentration parameter, we use the same time-adaptive
    training approach as for the climatological model; hence, separate models are
    estimated over all observation dates, using the same time of $31$~days
    centered around the day of interest over all available years in the training
    data \citep{Lang+Lerch+Mayr:2019}.
  \item \emph{Circular regression tree:}
    For the circular regression tree introduced in Section~\ref{sec:circtree},
    all covariates provided in the learning data can be considered due to an
    intrinsic automatic variable selection performed in the tree estimation. The
    tree is built with the newly developed \textsf{R} package \pkg{circtree} employing the
    CTree algorithm \citep{Hothorn+Hornik+Zeileis:2006} using a minimal number of
    2000 observations in each terminal node (argument \code{minbucket}).
  \item \emph{Circular regression forest:} Following the description in
    Section~\ref{sec:circforest}, the circular regression forest used in this study is
    constructed based on $100$~individual trees employing the \textsf{R} package
    \pkg{circtree}. Each of these trees is again built by the CTree algorithm on a
    subsample containing $30$~percent of the original learning data. All
    covariates are included for building each tree which ensures that the lagged
    response variable is always considered for splitting. This bagging approach can
    be applied in \pkg{circtree} by setting the argument \code{mtry} to the total
    number of covariates. Since a high number of possible split points leads to
    high computational costs, the covariates are binned in a maximum of $50$
    classes (argument \code{nmax = c(yx = Inf, z = 50)}). Contrary to a
    single-tree model, forests usually consist of very large trees as they are not
    prone to overfitting the data due to the stabilization obtained by combining
    the individual trees. Therefore, we use the following control arguments to
    build rather large trees: The minimal number of observations to perform a split
    is set to 20 (argument \texttt{minsplit}), the minimal number of observations
    in each terminal node is set to 7 (argument \texttt{minbucket}), and the
    significance level for variable selection is kept at its maximum value of $1$
    (argument \texttt{alpha}).
\end{itemize}

To compare the predictive performance of all proposed models, a circular
analogue of the continuous ranked probability score (CRPS) as introduced by
\cite{Grimit+Gneiting+Berrocal:2006} is computed. Just as the linear version of
the CRPS \citep[for more details see][]{Hersbach:2000} it is a proper scoring
rule \citep{Gneiting+Raftery:2007} and measures the difference between an
observation and the corresponding predicted distribution function in order to
assess the probabilistic goodness of fit for the estimated model. Hence, the
lower the CRPS value the better the predictive performance. Contrary to the
linear version, the circular CRPS reduces not to the absolute error but to the
angular distance when the forecast is deterministic.

In addition to the raw CRPS, corresponding skill scores are computed to assess
differences in the improvement of the various statistical models over the
climatological model used as a reference:
\begin{equation} 
  \text{CRPSS}_{\text{model}} = 1 -
  \frac{\text{CRPS}_{\text{model}}}{\text{CRPS}_{\text{climatology}}}.
\end{equation}

All scores presented in the next section are computed out-of-sample based on
five years of data. For the persistency model only dates prior to the time of
interest are used and the validation is performed rolling over all
observations. For all other models a five-fold cross-validation is employed
using up to four calendar years for model training and the remaining single
calendar year for validation. Due to the large sample size of 24 hourly values
per day over five years, some kind of temporal aggregation is needed to ensure
a correct visual comparison of the individual methods. The analyses performed
have shown that for the employed models the variability of the predictive
performance over the five years is lower than over a single day or over a
single year. Hence, CRPS and CRPS skill scores are aggregated over the
respective five validation years which yields 24 hourly scores per month
averaged over the five validation years.

\subsection{Results} 
\begin{figure}[t]
  \centering
  \setkeys{Gin}{width=1.0\textwidth}
  \includegraphics{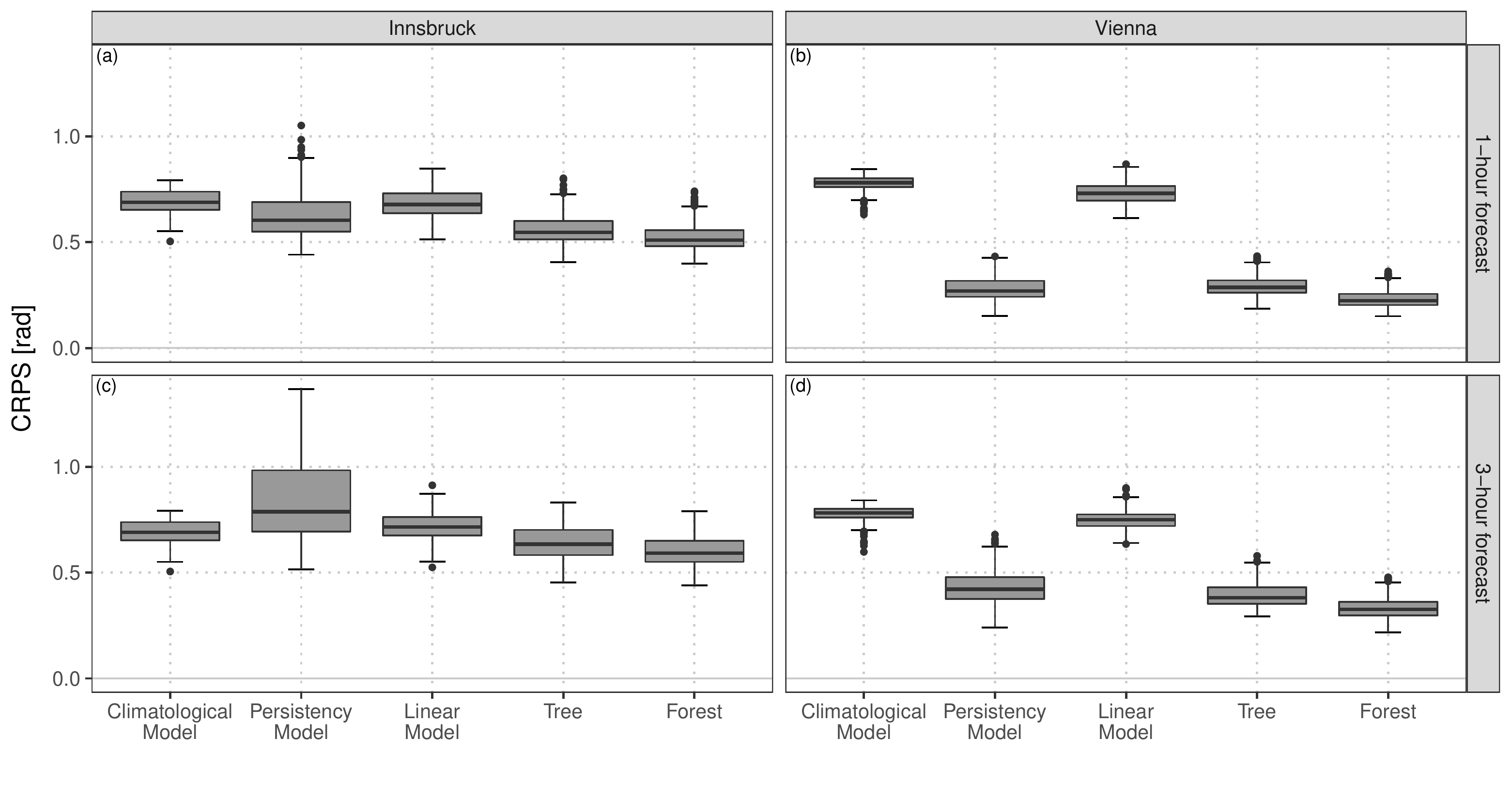}
  \caption{CRPS skill scores of wind direction forecasts based on the full
    predictive von~Mises distribution for $+1$\,h and $+3$\,h forecasts at the
    Innsbruck Airport and Vienna International Airport. Each box-and-whisker contains $24$~hourly scores
    for each of the $12$~months averaged over the $5$~validation years which
    yields a total of $288$~yearly mean values. The scores are shown for the
    climatological, persistency and the linear model as well as for the circular
    regression tree and forest.}
  \label{fig:boxplot_crpsraw} 
\end{figure}

This section provides a detailed analysis on the predictive performance of the
different proposed statistical models applied to probabilistic wind direction
forecasting. To ensure a comprehensive comparison of the models, wind direction
forecasts are evaluated for two different lead times at two airports with
different climatological site characteristics.
Figure~\ref{fig:boxplot_crpsraw} shows the CRPS values of the employed models
at forecast steps $+1$\,h and $+3$\,h for Innsbruck (Panels~a,\,c) and Vienna
(Panels~b,\,d). The scores are aggregated over the five validation years,
yielding yearly mean values for every hour per calendar month, with a lower
score indicating better performance. The circular regression forest overall
provides the best predictive performance, followed by the circular regression
tree and the persistency model for both stations at both forecast steps; except
for the $3$\,h~forecasts at Innsbruck where the persistency model is
outperformed by all others. In comparison to the circular regression tree and
forest, for both stations and forecast steps, the climatological model and the
linear model show clearly higher CRPS values and hence a lower predictive
performance.
The different site characteristics of the airports Innsbruck
(Figure~\ref{fig:boxplot_crpsraw}\,a,\,c) and Vienna
(Figure~\ref{fig:boxplot_crpsraw}\,b,\,d) seem to have an effect on the
absolute level of the model performances and on their respective predictive
performance variances. At Innsbruck, due to the surrounding mountains only a
limited number of possible wind directions exists, namely the three wind
regimes discussed for Figure~\ref{fig:tree_ibk} in
Section~\ref{sec:introduction}. Therefore, for Innsbruck the wind direction
remains rather constant in one of these possible states, but once a change
takes place it is mostly a major wind direction shift, e.g., from up-valley to
down-valley. Due to the few wind regimes the rather inflexible climatological
and linear models score relatively well with similar CRPS values as the other
models (Figure~\ref{fig:boxplot_crpsraw}\,a,\,c). In addition, at Innsbruck the
potential high prediction errors in case of a change of the wind regime seem to
lead to a higher variation in the predictive performance for all models in
comparison to Vienna; this variation is especially high for the persistency
model due to its strong vulnerability to abrupt wind shifts. On the contrary,
at Vienna smaller and less abrupt changes in the wind direction as well as less
pronounced wind regimes are observed due to the less mountainous surrounding.
This seems to weaken the predictive performance of the climatological and
linear models, and to reduce the performance variability for all models
(Figure~\ref{fig:boxplot_crpsraw}\,b,\,d).

The different forecast steps have apparently only a minor effect on the
predictive performance of the climatological model and the linear model at both
stations. As expected, for the persistency model, at both stations, higher
scores for the $3$\,h forecast (Figure~\ref{fig:boxplot_crpsraw}\,c,\,d) reveal
a lower performance for longer lead times; this is due to the lower information
content of 3-hourly instead of 1-hourly lagged response values employed as
covariates in the persistency model. The circular regression tree and forest
seem to partially compensate for the lower skill of the lagged response values
by other covariates, hence their predictive performance only slightly decreases
for the longer lead time. This compensation is especially evident for
Innsbruck, where the performance difference between the persistency model and
the tree-based methods significantly increases from the 1-hourly to the
3-hourly forecast.

\begin{figure}[t]
  \centering
  \setkeys{Gin}{width=1.0\textwidth}
  \includegraphics{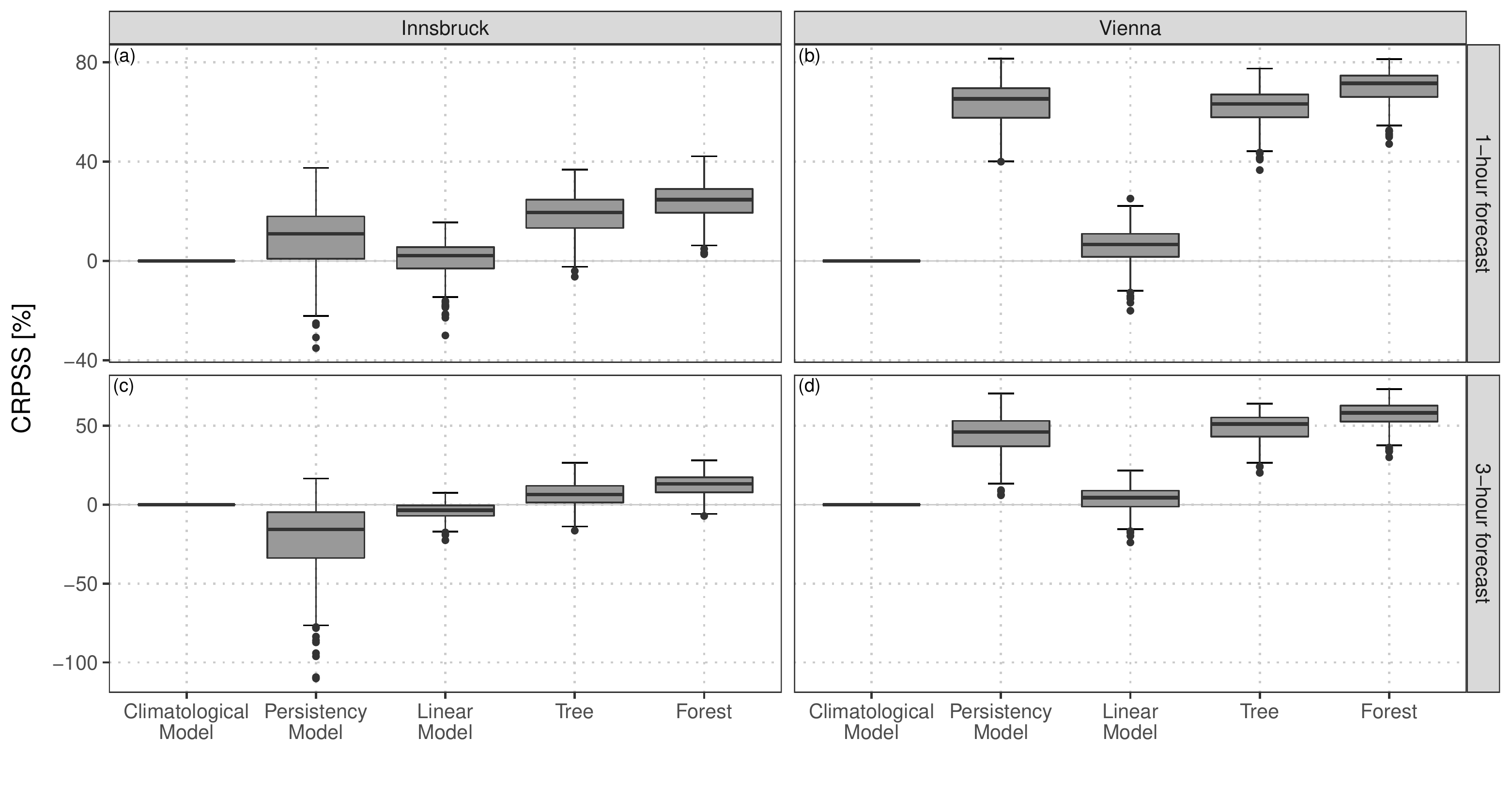}
  \caption{As Figure~\ref{fig:boxplot_crpsraw}, but showing CRPS skill scores
    with the climatology model as reference. Skill scores are in percent; positive
    values indicate improvements over the reference.} 
  \label{fig:boxplot_crpsskill} 
\end{figure}

In addition to the raw CRPS (Figure~\ref{fig:boxplot_crpsraw}), CRPS skill
scores with the climatological model as a reference are provided in
Figure~\ref{fig:boxplot_crpsskill}. Skill scores are in percent, where positive
values indicate an improvement in the predictive performance over the
reference. For all setups, the circular regression forest has the highest skill
scores with a mean performance gain of 13--25\% and 58--71\% for Innsbruck and
Vienna, respectively. As discussed for Figure~\ref{fig:boxplot_crpsraw}, this
improvement over the climatological model is lower for Innsbruck due to the low
number of predominant wind regimes and hence a relatively good performance of
the climatological model. Additionally, Figure~\ref{fig:boxplot_crpsskill}
shows that while the persistency model's performance is lower than the
reference (Panel~c) the tree-based models can compensate for the low skill of
the lagged response values employed as covariates and, hence, are still
significantly superior to the reference.

\section{Summary and conclusion}
\label{sec:summary}
Extending the toolbox for modeling circular data, circular regression trees and
forests are established by coupling model-based recursive partitioning with the
von~Mises distribution. By separating the data into more homogeneous subgroups,
possible difficulties in circular regression are bypassed as covariates are
solely considered for splitting and group-specific models are fitted without
further covariates. In addition, by specifying the von~Mises distribution for
each node and allowing for splits in both distribution parameters $\mu$ and
$\kappa$, fully probabilistic forecasts are provided.

The performance of the novel circular regression trees and forests is assessed
in a case study for shortterm probabilistic wind direction forecasting at two
airports with different site characteristics. As benchmark models,
probabilistic climatology and persistency models, as well as a state-of-the-art
circular GLM-type model are evaluated based on proper scoring rules. In
summary, the circular regression trees and forests have the highest predictive
performance in this setting. For cases without changes in the wind regime,
lagged response values provide already highly skillful estimates leading to a
good performance of the persistency model as observed for shortterm wind
direction forecasts in this study. While in these cases the trees and forests
also benefit from the highly informative lagged response, they can compensate
for a lower information of this covariate by incorporating other quantities and
possible interactions of these, contrary to the persistency model (see
Figure~\ref{fig:boxplot_crpsskill}). Hence, the tree-based models provide
reliable forecasts in all tested meteorological settings. For operational use,
a possible extension could be the incorporation of numerical weather
predictions as (additional) covariates. While this probably only slightly
improves the predictive skill for short leadtimes, it possibly extends the
potential forecast range of the different methods.

For the specific task of wind forecasting, the wind direction is often only
relevant if the wind speed is sufficiently high. Hence, it is of interest to
account for both quantities simultaneously, e.g., by considering a bivariate
normal distribution for wind vectors (from which wind speed and wind direction
can be obtained). The parameters of this bivariate normal distribution could
then be linked to available covariates using an additive regression framework
\citep[as proposed by][]{Lang+Mayr+Stauffer:2019} or using a tree-based
approach, similar to the one proposed in this paper. Moreover, a rather
different approach for a combined response of wind speed and wind direction
would be a two-step or hurdle model: In the first step this could build on the
truncated normal model of \cite{Thorarinsdottir+Gneiting:2010} to capture wind
speed; in the second step a circular wind direction model is leveraged given
that a certain hurdle for the wind speed is crossed.

Another possible improvement for obtaining more parsimonious circular
regression trees is to consider splitting circular covariates into two circle
segments by searching two split points simultaneously rather than sequentially
at different depths. While this might slightly improve the predictive
performance of circular regression trees, this should not affect the
performance of the forests, as they consist of very large trees with many
different splits.

To conclude, in general the tree structure can capture nonlinear changes,
shifts, and potential interactions in covariates without prespecification of
such effects. As supported by the presented case study, this can be
particularly useful for modeling a highly fluctuating response, such as
typically observed for wind direction, or/and in case of a large number of
possible covariates. Moreover, the case study shows that building ensembles of
circular regression trees can even improve the forecasting performance, as the
resulting forests allow for modeling smooth effects and stabilize the model.

\section*{Acknowledgments}
This project was partially funded by the Austrian Research Promotion
Agency~(FFG) grant number~$858537$. Torsten Hothorn received funding from the
Swiss National Science Foundation, grant number~$200021\_184603$. Lisa
Schlosser received a PhD scholarship granted from the University of Innsbruck.

\section*{Computational details}
The corresponding implementation of the proposed methodology for circular
regression trees and forests is provided in the \textsf{R} package
\textbf{circtree} (version~0.1.0). The package is
based on the \textbf{disttree} package
(version~0.2.0) which applies the main tree
building functions from the \textbf{partykit} package
(version~1.2.6). All three packages are available
on \textsf{R}-Forge at {\url{https://R-Forge.R-project.org/projects/partykit/}}.

For the circular GLM considered as reference model the corresponding
implementation is provided in the \textsf{R} package \textbf{circglmbayes} by
\cite{Mulder+Klugkist:2017}. In particular the function \code{circGLM} is
applied to estimates the intercept and regression coefficient along with the
concentration parameter.

\bibliography{ref.bib}

\newpage
\begin{appendix}

\section{Tree algorithm}
\label{app:algorithm}
This section provides a more detailed overview on the permutation-test-based
CTree algorithm \citep{Hothorn+Hornik+Zeileis:2006}, specifically for circular
data as applied for building circular regression trees and forests presented in
this case study. An alternative tree-building framework is provided by the MOB
algorithm, which is based on M-fluctuation tests \citep[see][for more
details]{Zeileis+Hothorn+Hornik:2008}.

In the following, the testing and splitting strategy is described for the root
node of the tree which contains the entire learning sample. For a complete
tree model, the same procedure is applied iteratively to all resulting child
nodes with the corresponding subsamples.

First, employing the von~Mises distribution, a distributional model
$\text{vM}(y; \mu,\kappa)$ is fitted to the learning sample of circular
observations $\{y_i\}_{i = 1,\ldots,n}$ as explained in
Section~\ref{sec:prob_circ}. In a next step, a goodness-of-fit measurement is
obtained for each parameter and each observation by evaluating the score
function $s(\mu, \kappa, y)$ at the estimated location and concentration
parameter~$\hat{\mu}$ and $\hat{\kappa}$. To detect dependencies between the
resulting scoring matrix
\begin{equation}
\begin{pmatrix} 
s(\hat{\mu}, \hat{\kappa}, y_1)_1 \quad, & s(\hat{\mu}, \hat{\kappa}, y_1)_2\\
\vdots & \vdots\\
s(\hat{\mu}, \hat{\kappa}, y_n)_1 \quad,& s(\hat{\mu}, \hat{\kappa}, y_n)_2
\end{pmatrix}
\end{equation}
and each possible split variable $z_l \in \{z_1, \ldots, z_m\}$ a permutation
test is applied. In particular, the null hypotheses of independence of each
split variable and the scores is assessed by employing the multivariate linear
statistic
\begin{equation}
t_l = vec\left(\sum_{i=1}^n v_l(z_{li}) \cdot s(\hat{\mu}, \hat{\kappa}, y_i)\right)
\end{equation}
with $s(\hat{\mu}, \hat{\kappa}, y_i) \in \mathbb{R}^{1\times 2}$. For a
numeric split variable $z_l$ the transformation function $v_l$ is simply the
identity function $v_l(z_{li}) = z_{li}$ such that $t_l \in \mathbb{R}^2$ as
the ``$vec$'' operator converts the matrix of dimension $1 \times 2$ into a $2$
column vector. If $z_l$ is a categorical variable with $h$ categories then
$v_l(z_{li}) = (\I(z_{li} = 1), \ldots, \I(z_{li} = h))$, hence, $v_l$ returns
a unit vector of dimension $h$ where the entry $1$ indicates the category of
$z_{li}$. In this case the ``$vec$'' operator converts the $h \times 2$~matrix
into a $h \cdot 2$ column vector by column-wise combination such that $t_l \in
\mathbb{R}^{h \cdot 2}$. If there are any observations with missing values
these are not included in the calculation of $t_l$.

To map the multivariate linear statistic $t_l$ onto the real line a univariate
test statistic~$c$ is employed, for example in a quadratic form
\begin{equation}
c_{\text{quad}}(t_l,\mu_l,\Sigma_l) = (t_l-\mu_l)\Sigma_l^+(t_l-\mu_l)^{\top}
\end{equation}
where $\mu_l$ and $\Sigma_l$ are the conditional expectation and the covariance
of $t_l$, as derived by \cite{Strasser+Weber:1999} and used for
standardization, and $\Sigma_l^+$ is the Moore-Penrose inverse of $\Sigma_l$.
As an alternative, also a maximum form ($c_{\text{max}}$) can be considered
such that the maximum of the absolute values of the standardized linear
statistic is returned.

The asymptotic conditional distribution of $c(t_{l},\mu_{l},\Sigma_{l})$ is
either normal (for $c_{\text{max}}$) or $\chi^2$ (for $c_{\text{quad}}$) owing
to the asymptotic conditional distribution of the linear statistic $t_l$ being
a multivariate normal with parameters $\mu_l$ and $\Sigma_l$
\citep{Strasser+Weber:1999}. With this knowledge at hand, the corresponding
$p$-values can be calculated and used to select the best splitting variable. A
small $p$-value corresponding to $c(t_{l},\mu_{l},\Sigma_{l})$ indicates a
strong discrepancy from the assumption of independence between the scores and
the split variable $z_l$. Therefore, if any of the Bonferroni-adjusted
$p$-values is beneath the selected significance level, the partitioning
variable $z_{l^\ast}$ with the lowest $p$-value is selected as split variable,
otherwise no split is performed. This early stopping induced by the
significance level is referred to as ``pre-pruning'' which is often avoided for
forest models by setting the significance level to $1$.

To select the best split point within the already chosen split variable, again,
a linear test statistic is employed. In particular, for a breakpoint $r$ of the
variable $z_{l^{\ast}}$ leading to two subgroups $\mathcal{B}_{1r}$ and
$\mathcal{B}_{2r}$ the discrepancy between score functions in the subgroups is
measured by evaluating
\begin{equation}
t_{l^{\ast}}^{qr} = \sum_{i \in \mathcal{B}_{qr}} s(\hat{\mu}, \hat{\kappa}, y_i)
\end{equation}
for $q \in \{1,2\}$. The breakpoint that leads to the highest discrepancy is
then selected as split point as defined by
\begin{equation}
r^{\ast} = \argmin_{r} (\min_{q=1,2}(c(t_{l^{\ast}}^{qr},\mu_{l^{\ast}}^{qr},\Sigma_{l^{\ast}}^{qr})).
\end{equation}

Subsequently, the same testing and splitting procedure is repeated in each of
the resulting subgroups until some stopping criterion is reached. Next to the
already mentioned significance-level-based stopping criterion, i.e., a minimal
$p$-value for the statistical tests, also a maximal tree depth or a minimal
number of observations in a node can be employed as stopping criteria.

\end{appendix}

\end{document}